# Visualization of association graphs for assisting the interpretation of classifications

SANJUAN ERIC[1]
ROCHE IVANA[2]

1 LITA, Université de Metz & SRDI, INIST – CNRS
IUT de Metz, Ile du Saulcy, 57045 Metz, CEDEX 1, FRANCE
2 SRDI, INIST – CNRS
2, Allée du Parc de Brabois, CS 10310, 54519 Vandoeuvre-lès-Nancy, FRANCE

## *Abstract*

Given a query on the PASCAL database maintained by the INIST, we design user interfaces to visualize and browse two types of graphs extracted from abstracts: 1) the graph of all associations between authors (co-author graph), 2) the graph of strong associations between authors and terms automatically extracted from abstracts and grouped using linguistic variations. We adapt for this purpose the TermWatch system that comprises a term extractor, a relation identifier which yields the terminological network and a clustering module. The results are output on two interfaces: a graphic one mapping the clusters in a 2D space and a terminological hypertext network allowing the user to interactively explore results and return to source texts.

## *1. Introduction*

As pointed out by Ley M. et al. (2006), traditional bibliographic data bases like PASCAL evolved from a small specialized bibliography to a digital library covering most of scientific fields. The collection is maintained with massive human effort. On the long term this investment is only justified if data quality remains much higher than that of the search engine style collections. In this paper we show that co-author and terminological graphs of high quality can be very easily extracted from PASCAL database, visualized and browsed. In particular, we founded out that special problems of person names can be managed using simple heuristics. Moreover, we show that is possible and useful to display the complete coauthor graph of several hundreds of PASCAL abstracts resulting from a request. This differs from Klink et al. (2004)'s approach that focus on local extracts of the graphs to comprehend only the surroundings of a single author on DBLP database.

We focus on valued graph of terms (words, names or key-words) that constitute the input to co-word analysis as defined in (Michelet, 1988). It has been observed by Courtial (1990) that in this type of analysis applied to scientific and technical databases, short paths of strong associations can reveal potential new connections between separated fragments of the network and reveal innovative situations.

Following these observations, graph clustering methods that do not focus on homogeneous clusters, but highlight some heterogeneous clusters formed along a path of strong associations have been proposed. The problem of graph clustering is well studied in a large literature. We refer to Flake et al. (2004) for a complete review. The best known graph clustering algorithms attempt to optimize specific criteria such as k-median, minimum sum, minimum diameter, etc., meanwhile simple fast variants of single link clustering (SLC) give satisfactory results in Courtial (1990)'s approach of co-word analysis. One of these variants can be found in the SDOC system (Polanco et al. 1995) actually integrated to STANALYST (http://stanalyst.inist.fr/) on-line interface that gives access to the PASCAL and FRANCIS databases (http://www.inist.fr) for information analysis purposes. It uses a threshold to fix the maximal cardinality of clusters. One of the most important qualities of SDOC is to present the content of clusters as subgraphs of associations that can be analyzed in a very intuitive way that minimizes the risk of observing correlations where they do not exists.

In this paper we revisit this main idea of co-word analysis applied to two other types of graphs extracted from bibliometric data: 1) the graph of all associations between authors, 2) the graph of strong associations between authors and terms automatically extracted from abstracts and grouped using linguistic variations. We adapt for this purpose the TermWatch system that comprises three main modules: a term extractor, a relation identifier which yields the terminological network and a clustering module. The results are output on two interfaces: a graphic one mapping the clusters in a 2D space and a terminological hypertext network allowing the user to interactively explore results, return to source texts or re-execute the system's modules.

Yet, based on the interactive graph visualisation toolkit (AiSee, http://www.aisee.com) and a different clustering algorithm called CPCL introduced in (Ibekwe-SanJuan 1995) and implemented in TermWatch (SanJuan 2005), we show how to reduce and explore informative paths between nodes.

The article is organized as follows. Section 2 is devoted to a precise description of our methodology which involves the definition of our text data representation §2.1, the graphs we extract from this data (§2.2 to §2.4) and the way we mine informative short paths §2.5 using a simple clustering algorithm. In §3 we carry out an experimentation of our approach on a small and a medium size corpus (the whole extracted graph feats in RAM memory). We process to the clustering of this non SWG in §3.3. §4 is devoted to discussion and future work.

## 2. *Graph extraction from documents*

### 2.1. *Association graphs*

We carry out the usual extraction of co-author and co-terms association graph from the collection of separate documents. From a formal point of view, the input is a hypergraph (a finite family *H*={*h1,h2, ...*} of finite subsets called edges) having as many hyper-edges as documents. Each hyper-edge *hi* is a set of units extracted from a document *i*. *hi* can contain authors, terms, bibliometric attributes etc.

From H_D, we derive the valued graph of associations *G*=(*V,E,a*) where:

- *V* is the set union of all extracted items.
- *E* is the family of all pairs (dyads or edges) of elements included in hyper-edges (for any edge *e* in *E*, *e* is of the form {*u,v*} and there exists an edge *di* containing both *u* and *v*).
- *a* is a valuation of *E*. In this experimentation we chose the equivalence coefficient which is defined for any edge *e={u,v}* as the product of conditional probabilities of finding one of the vertex in an hyper-edge knowing the presence of the other.

### 2.2. *Reducing and Visualizing Association Graphs*

Usually, when a co-occurrence is used, a threshold is set on the keyword frequency in order to obtain a less sparse matrix (Feldman et al., 1998). In our approach we prefer to set the threshold on association index after observing that low frequency items from PASCAL can represent valuable prospective rare information without inducing too much noise. Consequently, every value s in ]0,1[ induces a sub-graph Gs=(V,Es,a) where Es is the set of pairs of vertices {i,j} such that a(i,j)> s.

We use a variant of the single link clustering (SLC), called CPCL, originally introduced in Ibekwe-SanJuan (1998) and described hereafter, to form clusters of keywords related by geodesic paths made of relatively high associations. However, any variant of SLC that reduces its chain effect can produce interesting results in this context. In this experiment we use CPCL instead of SDOC because it does not require fixing the maximal size of a cluster, and relations between extracted clusters are symmetric. For all these reasons, CPCL is better adapted to the task of valuate graph reduction while preserving its structure. We use two types of interfaces to browse the resulting clustered network.

The interactive graph visualization interface AiSee (http://www.aisee.com) is based on minimization energy models. It displays the graph on the form of a set of masses related by spiral springs, represented by straight edges. Vertices represent units that can be interactively folded or clusters that can be unfolded to observe their internal structure and their external links to surround clusters or external items. Since clusters are labeled by their central terms, this interface reveals units (authors or terms) with high scores of betweenness centrality. Moreover, by opening the clusters the user visualize the geodesic paths of strong associations that cross the label of the cluster. These short paths reveal potential new interactions between document attributes.

In the case of very large graphs, like graphs including author's terminology as explained hereafter, a supplementary hypertext browsing interface is required to retrieve and explore informative associations. We used the navigator interface included in the TermWatch system to check details of association between authors and automatic extracted terms from raw text.

## *2.3. Clustering Algorithm*

There exist many graph clustering algorithms. In particular, Matsuda et al. (1999) proposed algorithms extracting high density subsets of vertices that could be apply to the values of graph G, but not to Gs graphs of textual data that are especially sparse. We chose the CPCL (Classification by Preferential Clustered Link) algorithm that tends to form clusters along short geodesic paths of strong associations. It consists in merging iteratively clusters of keywords related by an association strongest than any other in the external neighborhood. In other words, it works on local maximal edges instead of absolute maximal values like in standard SLC. We briefly recall this algorithm hereafter and we refer the reader to (Berry et al. 2004) for a detailed description in the graph formalism.

Program CPCL(V,E,a)

1) Compute the set S of edges {i,j} such that

a(i,j) is greater than s(i,z)

and s(j,z) for any vertex z,

2) Compute the set C of connected components of the sub-graph (V, S).

3) Compute the reduced valued graph (C, E_C, a_C)

where E_C is the set of pairs of components {I,J}
such that:

there exists {i,j} in E with i in J, j in J

and a_C(I,J) = max{a(i,j): i in I, j in J}.

If V <> C go to phase 1 else return (C, E_C, a_C)

## *2.4. Term extraction for topic mapping*

Termwatch performs multi-word term extraction based on shallow NLP, using the LTPOS tagger and LTChunker software (C) Andrei Mikheev 2000 of the University of Edinburgh. LTPOS is a probabilistic part-of-speech tagger based on Hidden Markov Models. LTChunk identifies simplex noun phrases (Nps), i.e., NPs without prepositional attachments. In order to extract more complex terms, contextual rules are used to identify complex terminological Nps.

Terms undergo variations which lead to the creation of new neighboring concepts. This process, called variation in the computational terminology field has been well studied (Ibekwe 1998). Variations occur at different linguistic levels: morphological, lexical, syntactic, semantic. TermWatch identifies the different variants of the same term going from close semantic variants like morphological (gender and spelling), synonyms (using WordNet) to generic-specific relations using syntactic criteria (expansions and structural changes).

Then given a collection $D=\{d1,...,dn\}$ of documents, we consider the following hyper graph $T=\{t1,...,tn\}$ where each hyper edge *ti* contains all author names of document *di* with all terms extracted in document *di* and all their variants founded in the corpus of documents.

From this hyper-graph we derive an association graph $G=(V,E,f)$ as explained in subsection 2.1 but with a different valuation function f. Indeed we set for each edge $(u,v)$ to:

- 1 if *u* and *v* are terms and *u* is a variant of *v*,
- a$(u,v)$ otherwise.

The graph Gs for s=0.8 is then clustered using CPCL and visualized in AiSee while we use the hypertext TermWatch interface to browse all association links between clustered elements as we already explained.

## *2.5. The station of analysis STANALYST®*

The station of analysis STANALYST® is composed of a whole of modules allowing the search for information in the bibliographical data bases of the INIST, for their statistical, terminological and thematic analysis (Polanco et al. 2001).

The integration of these various modules within a common, accessible graphic user interface since a navigator HTTP is as follows: the RECEPTION is a static page HTML giving access to the application; the user declares his name and defines his password. The PROJECT makes it possible to define an environment of work, i.e. a repertory in which will be stored all the results concerning the project. The user is the owner and it also has the possibility of giving access to its project to the associated users.

Modules CORPUS, BIBLIOMETRICS, INDEXING and INFORMETRICS constitute the modules of work of this station of analysis of information. The module CORPUS manages the creation of corpus by execution of requests built by the user. The corpora can then be exported bound for the following modules. Module BIBLIOMETRICS manages the creation of descriptive statistical analyses. The module INDEXING makes it possible either to revise the indexing or to carry out an automatic indexing of the corpus, resting for that in tools allowing a terminological extraction on the basis of several terminological references. The result of the INDEXING module will be the input for classification set of themes. The module INFORMETRICS manages classification using non-supervised automatic classification methods: two programs are available. The first realizes a hierarchic classification based on the co-word analysis method and the second one applies the K-means axial method to obtain a non-hierarchical classification.

The modules use a whole of repertories of work containing the programs, scripts, parameters necessary to its operation, as well as the whole of the projects created by the users.

# *3. Results*

## *3.1. Corpus extracted from PASCAL database*

We extracted two experimental corpora from PASCAL database. A corpus on South-America nano-technologies that we shall refer by SAN for short, and a corpus on chordal graphs that we shall refer by CG.

The SAN corpus is constituted by 939 bibliographic references coming from PASCAL database. The query asks for scientific and technical papers related to nanotechnologies, published in the twelve last years (1994 to nowadays) and having at least one author affiliated to a South-American organization. The principal scientific domains covered by the corpus references are Physics with 68% of the references, Engineering Sciences with 16% and Chemistry with 11%. 85% of the bibliographic references have been published in the 6 last years. Almost all the documents (99.7%) are in English language. The 2,574 authors come from 1,984 affiliations located in 51 different countries. The five best represented are Brazil, USA, Argentina, France and Spain with, respectively, 40%, 12%, 9%, 6% and 4% of documents. The documents come from 180 journals published in 11 different countries. The three first are USA, The Netherlands and United Kingdom with, respectively, 35%, 31% and 16% of documents. The 4 journals the most represented (2% of the cover) produce 23% of the references and 75% of them are produced by 28% of the cover.

The CG corpus is composed by 155 bibliographic references coming from PASCAL database. The query asks for scientific papers dealing with chordal graphs and explores 22 years of PASCAL database. The principal scientific domains covered by the corpus references are Mathematics (56%) and Engineering Sciences (43%). 67% of the bibliographic references have been published in the 6 last years. All the documents are in English. The 261 authors come from 240 affiliations situated in 32 different countries. USA, Canada, France, Germany and Taiwan with, respectively, 27%, 10%, 9%, 9% and 4% of documents are the best represented. The documents come from 29 journals published in 8 different countries. The Netherlands, Germany and USA with, respectively, 57%, 29% and 29% are the first ones. 25% of the references are produced by only one journal and the five first journals produce 75% of the cover.

## *3.2. Co-Author graphs*

We generated two level visualizations of author graphs. First level is the whole graph of associations between authors, thus all pairs of authors that wrote at least one paper together are represented. For sets with less than 300 documents this graph can be browsed using an appropriate toolkit like AiSee as shown in Fig. 1 (left side) for the SAN corpus and Fig 3 for the corpus on chordal graph.

Fig 1. Co-author graph extracted from corpus SAN. Left figure shows the general shape of the graph. Vertices represent authors. Neighboring authors in the same cluster share the same color. Right figure shows the contents of clusters Knobel and Souza.

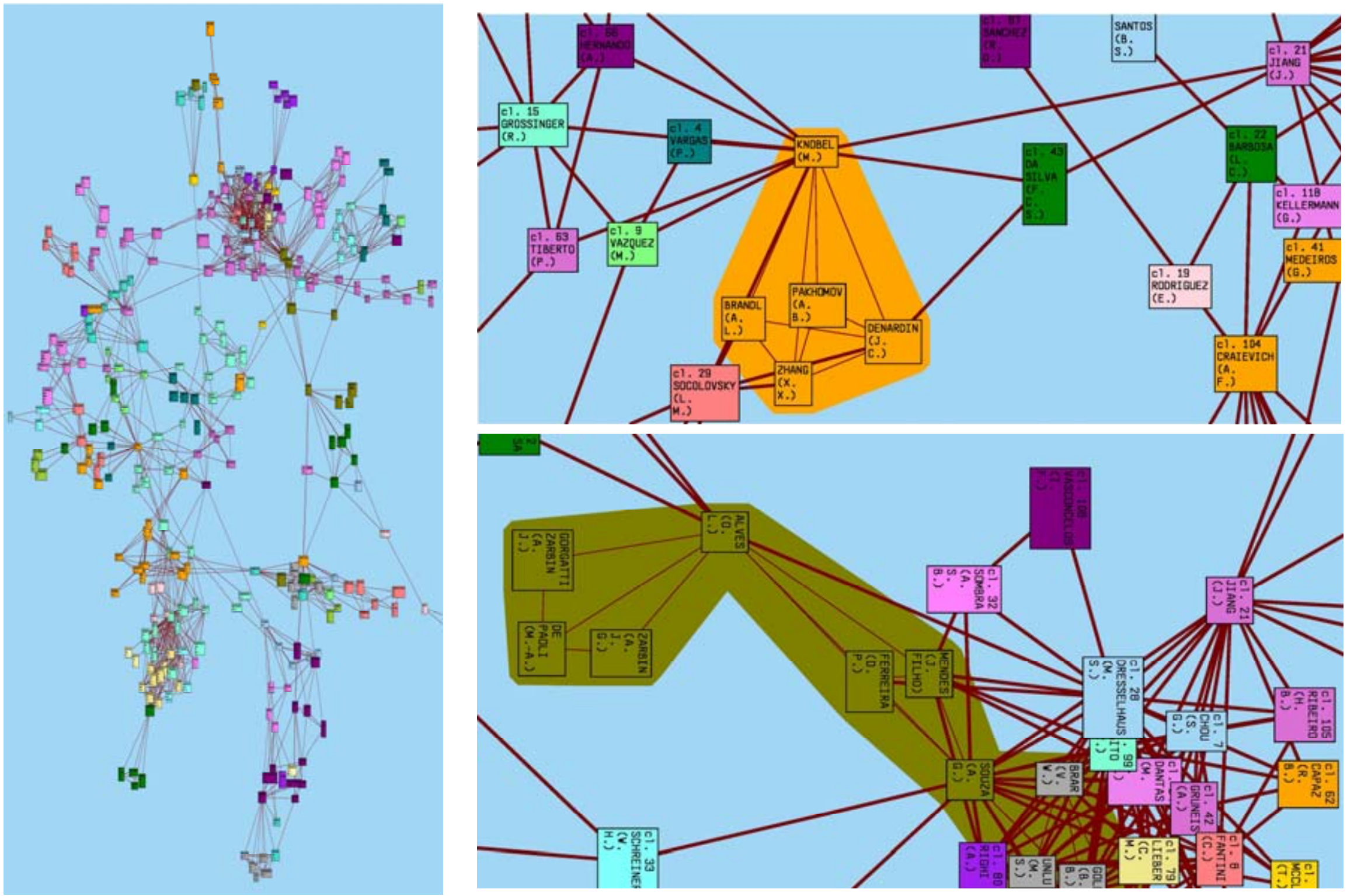

In the case of SAN corpus, co-author graph reveals a set of clusters representing international important collaborations of South-American academic institutions. It is possible to observe in cluster

J. Jiang a great number of exchanges with Japan and in cluster M. S. Dresselhaus a strong collaboration with USA. Some authors have a central position as M. Knobel, A. F. Craievich, P. C. Morais, A. G. Souza or D. Ugarte, respectively in clusters M. Knobel, A. F. Craievich, P. C. Morais, A. G. Souza and J. Jiang. These authors come from internationally known academic institutes. It is also possible to find author cliques: an interesting example can be observed in cluster P. Levy, whose authors are related to a national atomic energy institute.

Fig. 2 Reduced co-author graph on SAN. Vertex represent clusters labeled by their author having the highest number of links towards other clusters.

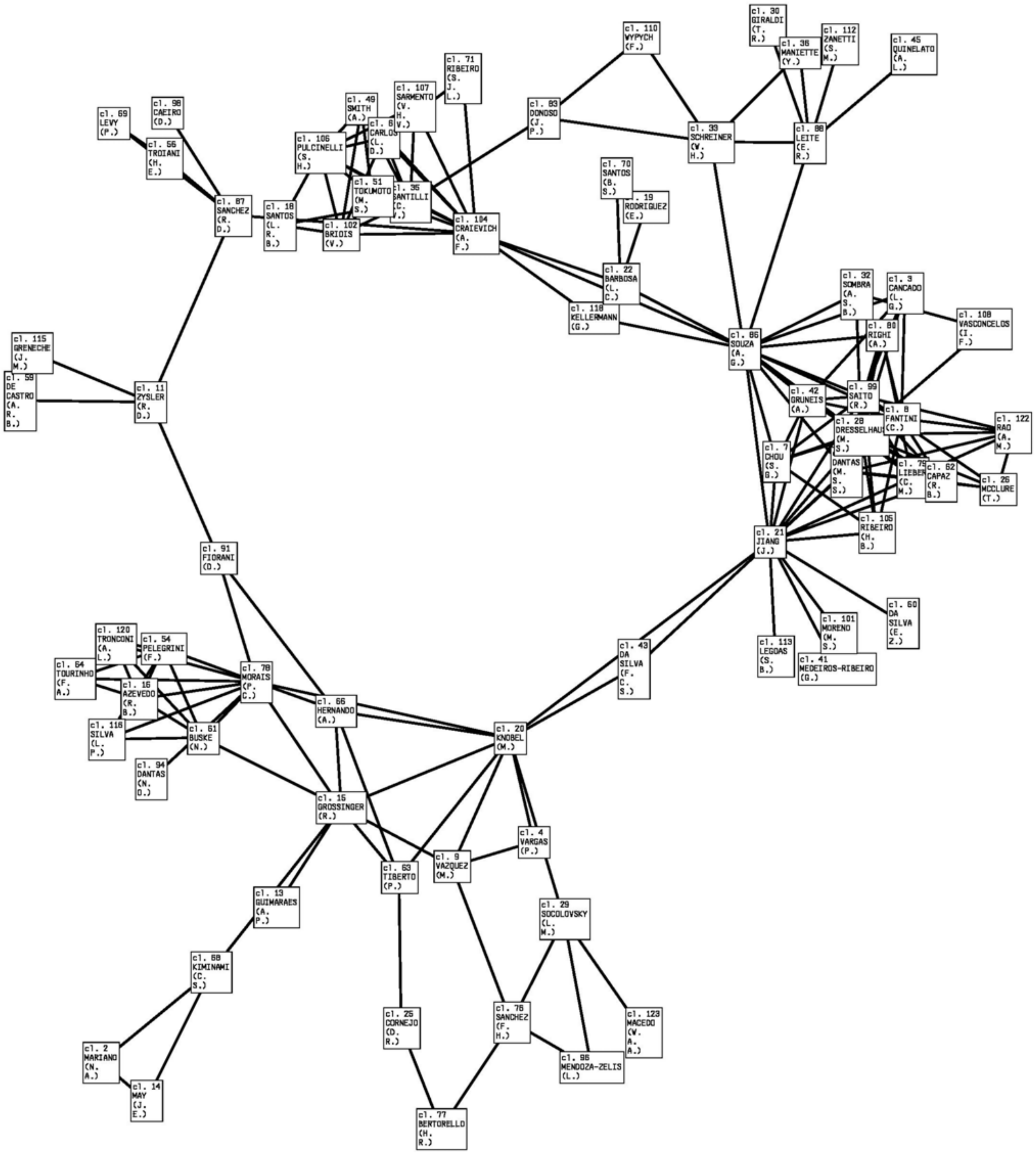

Clustering allows to highlight an underlying structure as shown in Fig 2 for the SAN corpus and authors that relate different groups as shown in Fig. 1 (right side).

In the case of CG corpus, co-author graph reveals a central author (cutting vertex): Dieter Kratsh and a dense cluster (unfolded in the figure) formed around Heggernes and Berry who is related to all authors in this cluster. Another feature of this image is to show the evolution of authors Berry and Paul since their PhD with Bordat and Habib respectively.

Fig. 3 Co-author graph extracted from CG corpus. Clusters are wrapped together.

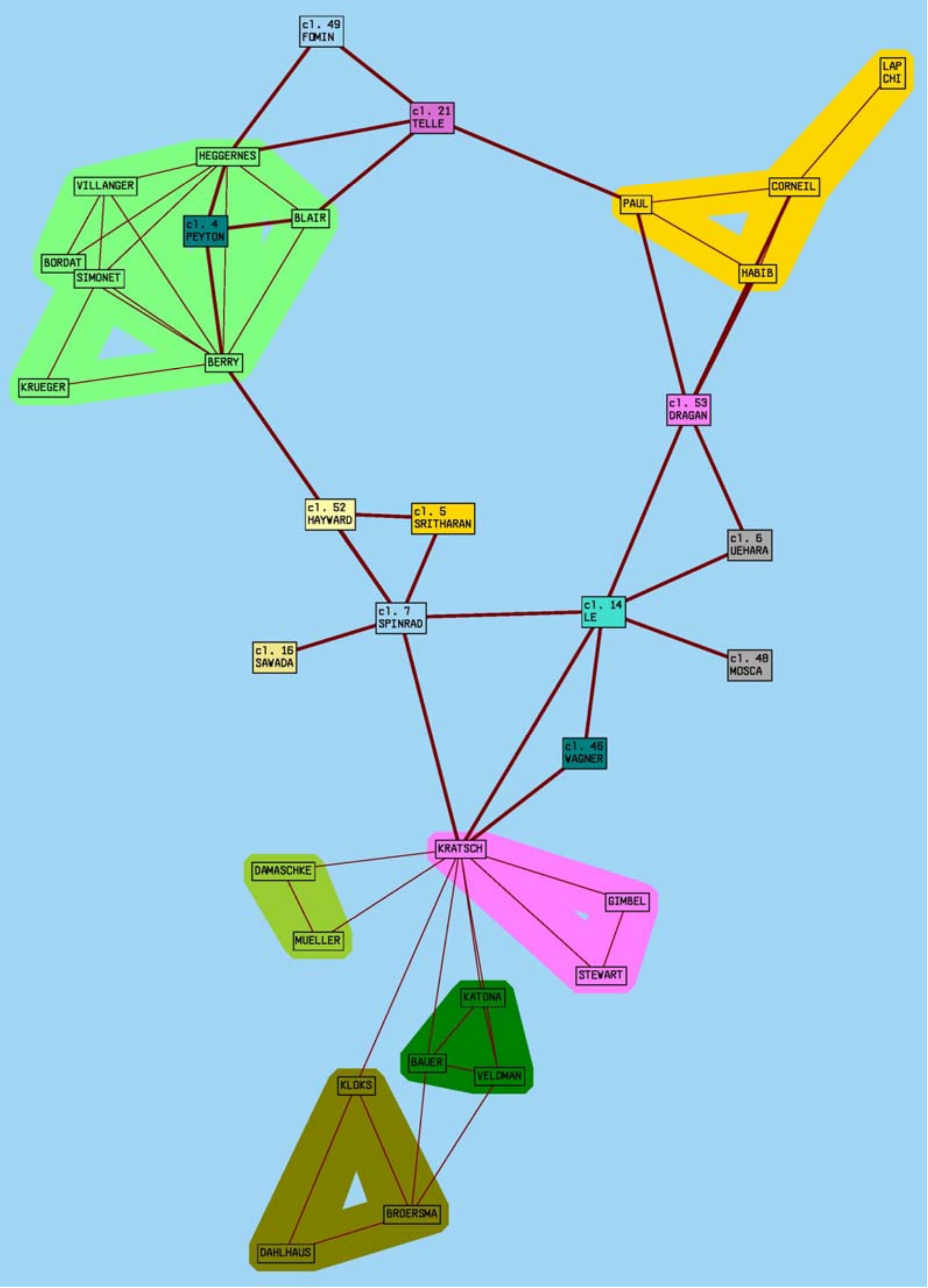

## *3.3. Terminology Author graphs*

We proceed to a term extraction in both corpora. The resulting graphs are huge and a hypertext browsing interface is required. However, in the case of GC corpus, graph display confirms the central position of Dieter Kratsch in this corpus since a cluster labeled by his name appears. By unfolding it, we see the terminology used in his publications.

Fig. 4 Cluster labeled Kratsch and associations to cluster of open questions

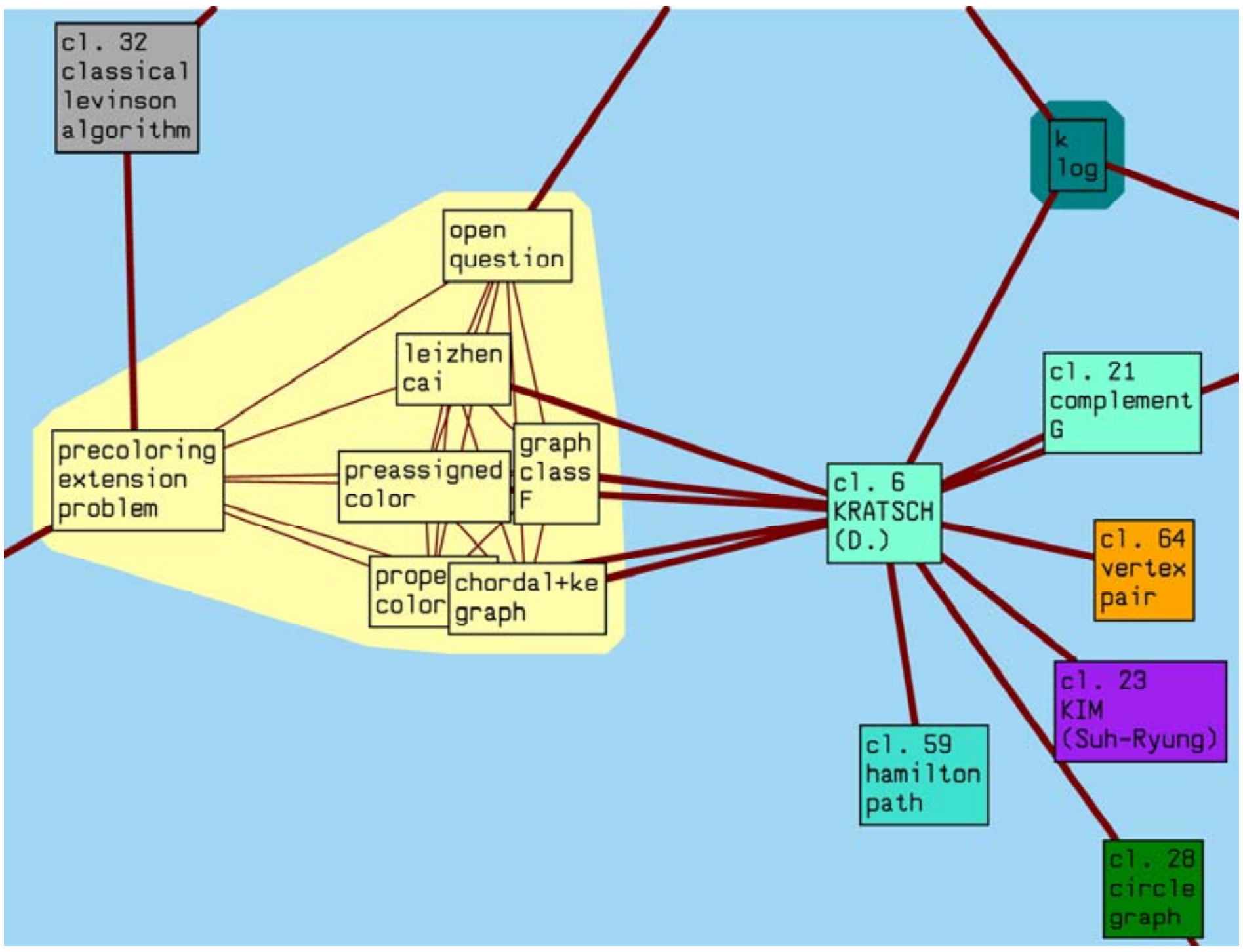

Fig. 5 Contents of cluster k log in GC graph.

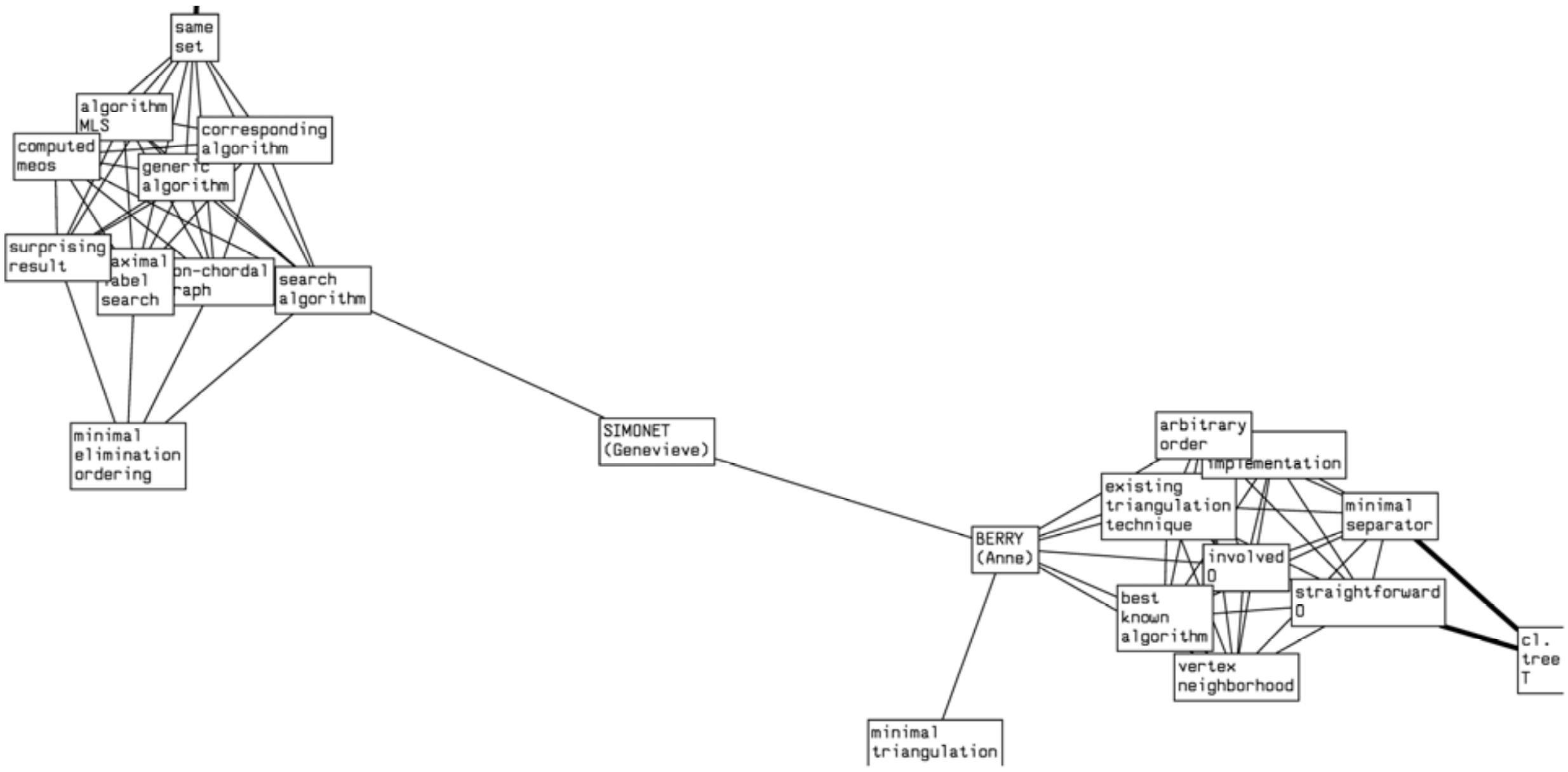

Fig 4 and 5 show features that is difficult to detect using the sole hypertext browsing interface. In the case of CG corpus, if the centrality of author Krastch is revealed by the fact that this name is the label of the biggest cluster, the graph display interface allows to point out the multiple links between terms associated to open questions and terms by folding the huge cluster labeled Krastch and unfolding the cluster labeled open question. By unfolding the two clusters of terminology respectively related to authors Simonet and Berry, that graphical interface allows pointing out the multiple graph problems targeted by Berry, meanwhile Simonet appears to be more specialized in a single application of these algorithms.

The terminology-author graph obtained from SAN corpus allows to confirm same statements made from the co-author graph exploitation. By example, it is possible to validate the central position of A. Craeivich and determine the terminology associated by examining the cluster contents: doped film, grazing-incidence small angle X-ray reflectivity, GISAXS pattern, film surface.

The cluster K. D. Machado presents the same clique of authors observed in the co-author graph (J. C. de Lima, T. A. Grandi, C. E. M. Campos and K. D. Machado) and the association with two others clusters: P. S. Pizani and binary mixture. The cluster binary mixture is composed by: hexagonal Co Se, nominal composition Co, amorphous selenium.

## *4. Discussion*

Relying on the data quality of PASCAL database, we have designed and experimented an interface that can extract and efficiently display in real time the co-author and terminology network from documents issued by a user query through the STANALYST system. Further human computer interface study shall allow us to better adapt this interface to non AiSee's users by automatically selecting clusters to fold or unfold for different scientific watch tasks. Moreover, in our experiment we did not use the key-word field of PASCAL abstracts since we focused on the feasibility of displaying the vocabulary used in text abstracts. Another further experiment will be to add this data to the hypergraph representation of documents.